\documentclass[a4paper,11pt]{article}
\usepackage{pos}

\title{Observation of VER J2019+368 with the SST-1M stereoscopic system}
 \ShortTitle{VER J2019+368 with SST-1M}

\author*[a]{J.~Jury\v{s}ek}
\author[a]{T.~Tavernier}
\author[a]{A.~L.~M\"{u}ller}

\scriptsize
\affiliation[]{
$^a$FZU - Institute of Physics of the Czech Academy of Sciences, Na Slovance 1999/2, Prague 8, Czech Republic,
}

\onbehalf{on behalf of the SST-1M Collaboration\hyperref[authorlist]{*}} 

\emailAdd{jurysek@fzu.cz}

\abstract{
The Single-Mirror Small Size Telescope  (SST-1M) is a small Cherenkov telescope designed to detect gamma rays with energies more than about 1 TeV. The optical design of the SST-1M follows the Davies-Cotton concept to ensure good off-axis performance. In 2022, two SST-1M telescope prototypes were installed in Ond\v{r}ejov, Czech Republic, and stereoscopic observations of astrophysical gamma-ray sources have been performed since then. VER J2019+368 is an unidentified very-high-energy (VHE) gamma-ray source, surrounded by several gamma-ray point-like and diffuse sources, together with their multi-wavelength counterparts. VHE emission was discovered by MILAGRO in 2012, followed by VERITAS observation, which revealed the complex morphology of the source. Recently, the LHAASO observatory detected photons with multi-TeV energies, opening up the possibility of particle acceleration up to PeV energies. In this contribution, we present preliminary results of the first observing campaign of the VER J2019+368 region, performed with SST-1M from April to November 2024. We present the data analysis, focusing on the morphological and spectroscopic study of the region. We also present the off-axis performance of SST-1M in the context of the prospects for detecting extended galactic gamma-ray sources. As one of the brightest and hardest sources in the LHAASO catalog, VER J2019+368 is an ideal candidate for testing the capabilities of the SST-1M, with its large field of view, to detect extended gamma-ray sources.
}

\ConferenceLogo{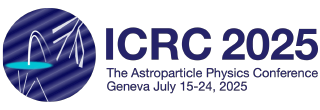}

\FullConference{39th International Cosmic Ray Conference (ICRC2025)\\
 15-24 July 2025\\
Geneva, Switzerland\\}

\begin{document}
\maketitle

%%%%%%%%%%%
\section{Introduction}

Surroundings of VER J2019+368 is a complex region with several high-energy (HE) to ultra-high-energy (UHE) gamma-ray point-like and diffuse sources accompanied by their multi-wavelength counterparts, potentially powering the very-high-energy (VHE) emission. There is HII region Sh 2-104, Wolf-Rayet star WR141, two pulsars detected by Fermi-LAT in HE gamma-rays, PSR J2021+3651 and PSR J2017+3625, the pulsar wind nebula (PWN) G75.1+0.2 detected in X-rays (the so-called "Dragonfly") \citep{2004ApJ...612..389H, 2008ApJ...680.1417V}, and the X-ray transient IGR J20188+3657 \citep{2014ApJ...788...78A}. MILAGRO discovered the VHE emission in 2012 \citep{Abdo_2012}, making it the second brightest MILAGRO source in the northern hemisphere. The region was later resolved into more VHE sources by VERITAS \citep{2014ApJ...788...78A, Abeysekara_2018}, revealing a complex nature and suggesting an energy-dependent morphology. Recently, the LHAASO observatory detected photons with energies up to 270 TeV, opening up the possibility of particle acceleration up to PeV energies \citep{2024ApJS..271...25C}. A recent multiwavelength study associated VER J2019+368 with the PWN around PSR J2021+3651 \citep{2023ApJ...954....9W}. Still, there is an open question if PSR J2021+3651, whose distance estimates differ significantly in the literature \citep{2015ApJ...802...17K}, can power the entire VHE emission.

As one of the brightest and hardest sources in the LHAASO catalog, VER J2019+368 is an ideal candidate for testing the capabilities of the Single-Mirror Small Size Telescope (SST-1M) with its large field of view (FoV), to detect extended sources, opening the door for further studies of Galactic PeVatron candidates. The SST-1M is a small Cherenkov telescope designed to detect gamma rays with energies greater than about 1 TeV \citep{sst1m_hw_paper, sst1m_performance_paper}. Two SST-1M telescope prototypes separated by 155.2 m were installed in Ond\v{r}ejov, Czech Republic (510 m a.s.l.), in 2022, and observations of astrophysical gamma-ray sources have been performed since then. The observations are conducted in stereoscopic regime with the event timestamps synchronized using the White Rabbit timing network \citep{white_rabbit}, and triggering managed with the CTA Software array trigger (SWAT).
%The SST-1M mirror dish (4 m in diameter) is composed of 18 hexagonal facets, and the optical design of the telescope follows the Davies-Cotton concept to ensure good off-axis performance. The camera consists of 1296 silicon photomultiplier (SiPM) pixels and a fully digital trigger and readout system \citep{2017EPJC...77...47H}. The SiPM technology allows high night sky background operation, significantly increasing the duty cycle compared to standard photomultipliers.

We present preliminary results of the first observing campaign of the VER J2019+368 region, performed with SST-1M from April to November 2024, resulting in 150 h of stereo data. We present the data analysis, focusing on the morphological and spectroscopic study of the region. We also discuss the off-axis performance of SST-1M, crucial for observations of extended sources, and follow-ups of
poorly localized transients.

%%%%%%%%%%%
\section{SST-1M off-axis performance}

Observations of Imaging Atmospheric Cherenkov telescopes (IACTs) are usually taken in the wobble mode \citep{1994APh.....2..137F}, where the telescope pointing direction has a small offset with respect to the position of the source. It allows for the estimation of the background from the same data, where the source is present in the FoV under the assumption of radially symmetric acceptance. Observation of extended sources presents a challenge for most existing IACTs due to requirements on the size of the FoV and the performance, which usually decreases rapidly with offset \citep[e.g.][]{2016APh....72...76A}. The good optical performance across the large optical FoV of SST-1M ($9^\circ$) ensured by the Davies-Cotton design of the optics, together with the triggering system, which includes all camera pixels, provides a unique off-axis performance for detection of gamma-ray sources \citep{2017EPJC...77...47H, sst1m_hw_paper, sst1m_performance_paper}. 

\begin{figure}[!t]
\centering
\begin{tabular}{c}
\includegraphics[width=.49\textwidth]{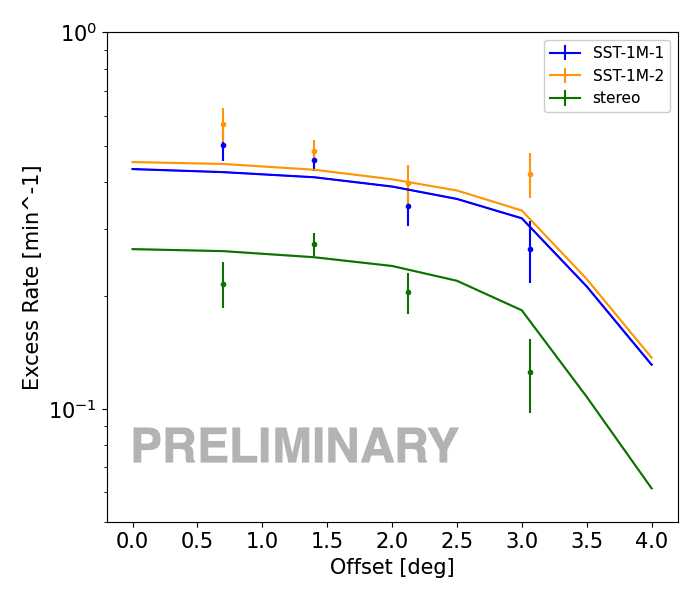} \\
\end{tabular}
\caption{Acceptance of SST-1M as a function of the offset angle. The lines represent the integral rate of simulated gammas with Crab Nebula spectrum in the signal region after analysis cuts as a function of the offset at $30^\circ$ zenith angle. The points show a measured rate of gamma-like excess events resulting from a preliminary analysis of the Crab Nebula observation.}
\label{fig.offaxis_performance}
\end{figure}

Figure~\ref{fig.offaxis_performance} shows the integral rate of simulated gammas with the Crab Nebula spectrum at $30^\circ$ zenith angle above the energy threshold. The events are integrated in the signal region, and only events surviving the analysis cuts are kept for both telescopes in mono and stereo. One can note that the gamma-ray acceptance is almost flat up to $\sim2.5^\circ$ where it drops by $\sim10\%$, which makes the SST-1M an ideal instrument for observation of extended gamma-ray sources, or poorly localized transients \citep{sst1m_performance_paper}. 

In order to validate the MC results, we performed dedicated observations of the Crab Nebula between $25^\circ$ and $35^\circ$ zenith angles at different wobble offsets. After the quality cuts, we collected the following amount of good-quality data for each offset: $0.7^\circ$: 6h, $1.4^\circ$: 19h, $2.1^\circ$: 7h, $3.06^\circ$: 4h (in both mono and stereo modes of data acquisition). The number of events surviving energy-dependent gammaness cuts in the signal and background regions above the energy threshold at the analysis level ($E_\mathrm{th} > 1.5$ TeV for mono, 2.0 TeV for stereo) was calculated. The result of a preliminary analysis is shown in Figure~\ref{fig.offaxis_performance}, confirming that even at $3^\circ$ offset, the drop of the gamma-ray acceptance is relatively small compared to the on-axis gammas.

%sst1mpipe_workdir/notebooks/mc_offaxis_perfomance.ipynb

%%%%%%%%%%%%%%%
\section{Observation and data analysis}

The first observation campaign of the VER J2019+368 region with the SST-1M took place from April to November 2024, resulting in a sample of 150 hours of data between $12^\circ$ and $60^\circ$ zenith angles. The observation was taken in the stereo mode with a wobble offset of $1.4^\circ$ with respect to the MILAGRO source coordinates. We applied the run selection based on monitoring of the rate of cosmic ray events \citep{sst1m_performance_paper} to remove runs affected by bad weather conditions or malfunctions of the instrument. The total amount of good stereoscopic data used in this analysis was 97 hours.

The raw pixel waveforms were calibrated and processed up to the photon list with reconstructed energies and arrival directions (see \citep{sst1m_performance_paper} for the details of the analysis and reconstruction) using \texttt{sst1mpipe}\footnote{\url{https://github.com/SST-1M-collaboration/sst1mpipe}} \citep{sst1mpipe_073}, which is the standard data processing and analysis tool for SST-1M. The Monte Carlo (MC) simulations were produced at fixed azimuth (south pointing) and zenith angles between $20^\circ$ and $60^\circ$ with a $10^\circ$ step. For the random forest (RF) reconstruction and to create the instrument response functions (IRFs), the closest node was selected for each observation run (about 20 minutes long).

The observations were collected over a relatively long period of time under variable atmospheric conditions on the site, while only a single MC production with fixed atmosphere was used for the RF reconstruction and to produce the IRFs for the preliminary analysis. To take into account the difference in the shower parameters between the MC and the real data, we scaled the Hillas intensities in the data by a ratio between the rate of the cosmic rays in MC and in the data. The ratio was calculated for events with Hillas intensities > 400 p.e. to avoid systematics induced by varying energy threshold. A preliminary study of systematic effects induced with such scaling showed that on top of the systematics described in \citep{sst1m_performance_paper}, one may expect additional $8\%$ and $5\%$ systematics on the flux normalization and the spectral index, respectively.

To select the candidate photons in the data, we applied an energy-dependent cut on gammaness reconstructed with the RF classifier \citep{2023arXiv230709799J, sst1m_performance_paper}. As the optimal cut can vary for different source fluxes and spectral energy distribution (SED), we optimized on MC, requesting the maximum detection significance for a source with $10\%$ of the Crab Nebula flux and the spectral index of $-2$, which resulted in the optimal cut keeping $50\%$ of the reconstructed MC gammas in each energy bin.

%%%%%%%%%%%%%%%
\section{Results}

% PSR J2021+3651
% DM 371 pc/cm3 https://arxiv.org/pdf/astro-ph/0206443 , they say it's huge and the derived distance is 19 pc
% Using 
% https://apps.datacentral.org.au/pygedm/
% https://www.cambridge.org/core/services/aop-cambridge-core/content/view/D8B2ABEBBB77F5D07BDE5E5CC3FE164B/S1323358021000333a.pdf/a-comparison-of-galactic-electron-density-models-using-pygedm.pdf
% the distance (for that DM) is 12.39 (NE2001) or 10.63 (YWM16)

Resulting photon lists combined with IRFs were processed in \texttt{gammapy} \citep{axel_donath_2021_5721467}. We produced a skymap of the region to address the morphology of VHE sources, and to find the signal regions for the spectral analysis.  

\subsection{Morphology of the source} \label{sec.morphology}
\begin{figure}[!t]
\centering
\begin{tabular}{cc}
\includegraphics[width=.49\textwidth]{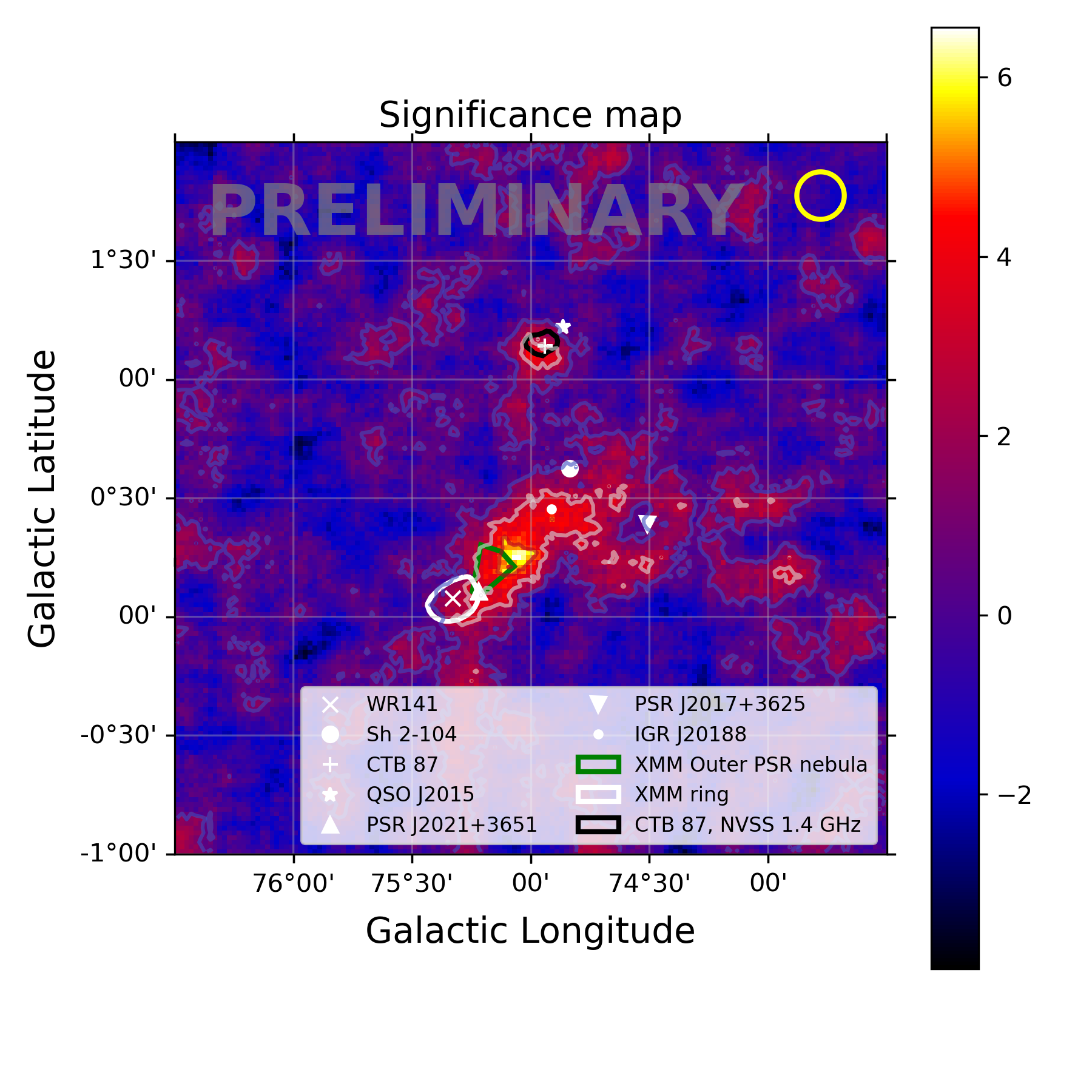} & \includegraphics[width=.49\textwidth]{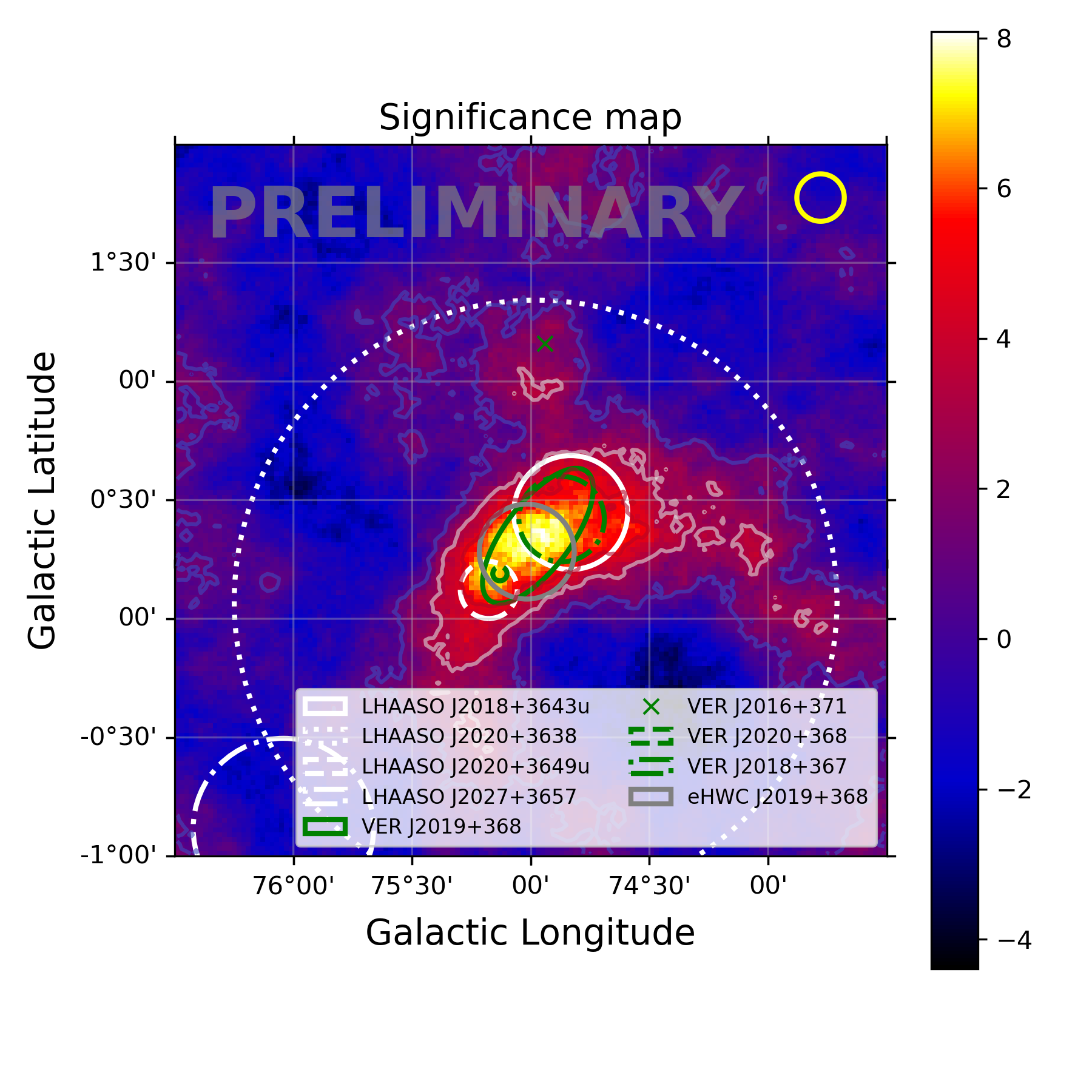} \\
\end{tabular}
\caption{Maps of local significance for the VER J2019+368 region as seen with the SST-1M stereoscopic system. Point-like ($0.1^\circ$, \textit{left}) and extended ($0.25^\circ$, \textit{right}) integration radii were used to produce the skymaps. \textit{Left} skymap shows the multi-wavelength counterparts, and \textit{right} skymap shows the VHE sources in the region. The radii of the VHE sources correspond to their extensions as reported in \citep{Abeysekara_2018, 2020PhRvL.124b1102A, 2024ApJS..271...25C}. The contours in both skymaps represent $1\sigma$, $3\sigma$, and $5\sigma$ local significance. The yellow circle in the top right corner of both skymaps represents the integrated PSF for the energy range shown in the map.}
\label{fig.dragonfly_skymap}
\end{figure}

Figure~\ref{fig.dragonfly_skymap} shows $3^\circ \times 3^\circ$ local significance maps of the VER J2019+368 region obtained with the ring background method in \texttt{gammapy} \citep{axel_donath_2021_5721467}. Regions within a radius of $0.3^\circ$ from all known nearby VHE sources were removed for background estimation. The energy range for the map is $1-300$ TeV with a safe mask requiring the effective area >1\% of the maximum, and energy bias <30\%. The convolution kernel used in left Figure~\ref{fig.dragonfly_skymap} is 0.1 deg, corresponding to the SST-1M stereo point spread function (PSF). This analysis is sensitive to the detection of point-like sources. We also tested for the presence of extended sources by convolution of the skymap with a kernel of 0.25 deg radius (Figure~\ref{fig.dragonfly_skymap}, right). The main excess was detected with $8.1\sigma$ pre-trial local significance using the extended convolution kernel, and its position corresponds well with the VHE sources detected by other observatories, which are marked together with their extension in Figure~\ref{fig.dragonfly_skymap}.

We checked the distribution of local significance for the background pixels outside the exclusion regions and confirmed that they are well Gaussian, with a mean of -0.09 and a standard deviation of 1.06. We performed two tests to check the skymap for possible artifacts due to a potential bias in the background estimation. First, we repeated the ring background estimation, varying the radius and width of the ring. As a second test, we run a completely independent analysis with the background model using the \texttt{BAccMod}\footnote{\url{https://github.com/mdebony/BAccMod/tree/main}} tool. Both tests provided consistent results.

To obtain the post-trial significance of individual excesses, we run dedicated MC simulations of the background events with IRFs for 30 deg zenith angle, resulting in 3,000,000 skymaps, processed the very same way as the collected dataset. The local significances were collected for all simulated skymaps, and a global probability of excess with a given local significance was calculated. The resulting post-trial significance is $4.8\sigma$ and $6.5\sigma$ for the VER J2019+368 region in point-like and extended source searches, respectively. Post-trial significance of the point-like source VER J2016+371, corresponding to the Supernova Remnant CTB87, is $1.7\sigma$. 

%sst1mpipe_workdir/data_pipe/notebooks/sst1m_significance_map_dragonfly_v074.ipynb

\begin{figure}[!t]
\centering
\begin{tabular}{cc}
\includegraphics[width=.48\textwidth]{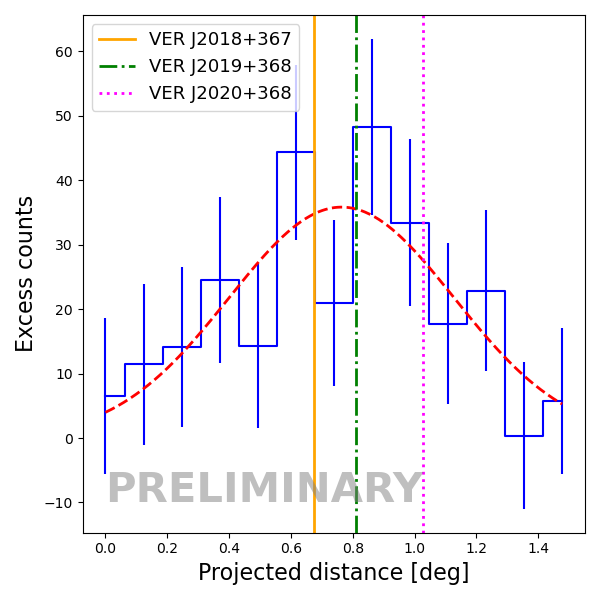} &
\includegraphics[width=.52\textwidth]{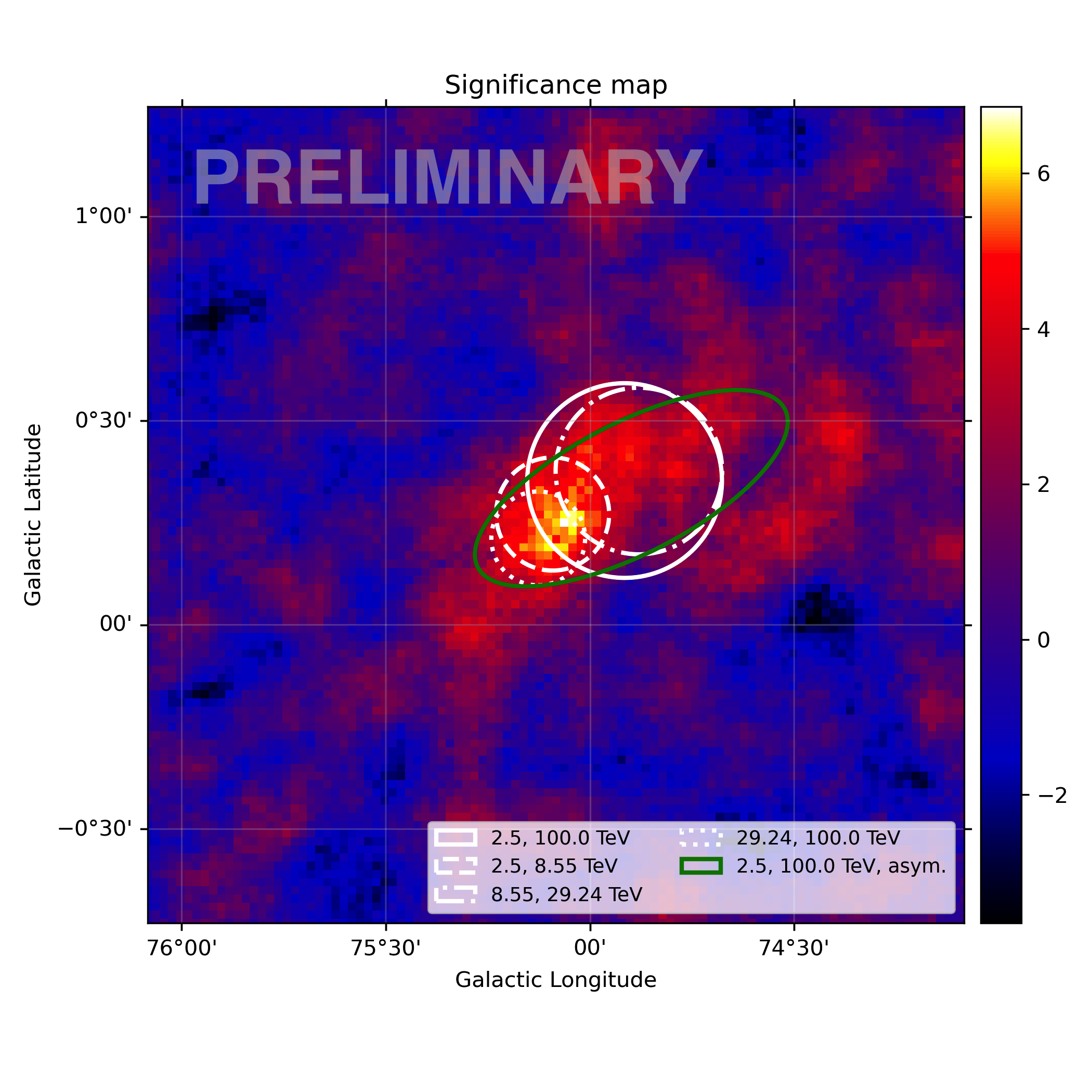} \\
\end{tabular}
\caption{\textit{Left:} The profile of VER J2019+368 emission along its main axis. The excess counts (blue distribution) are calculated from the skymap constructed using the point-like integration region. The red line represents the best-fitting Gaussian, and the vertical lines mark the coordinates of the three VERITAS sources along the profile. \textit{Right:} The map of local significance between 2.5 TeV and 100 TeV showing the best-fitting source morphologies. The white lines represent the extension of a symmetrical Gaussian in different energy ranges. The green line shows the best-fitting asymmetric Gaussian on the full energy range.}
\label{fig.dragonfly_profile}
\end{figure}

The main VER J2019+368 region has an asymmetric morphology. We created a profile of the excess along the main axis of the source in the point-like search skymap, where each bin was produced using an aperture of width $0.12^\circ$ and height $0.3^\circ$ (Figure~\ref{fig.dragonfly_profile}, left). To assess the source extension, we fit the profile using a single Gaussian with a standard deviation of $0.36\pm0.08$, which is consistent with the extension reported by VERITAS \citep{Abeysekara_2018}. We note, however, that we cannot confirm the two-source morphology (VER J2018+367 and VER J2020+368) suggested by the same authors.

%We report a hint of an additional point-like emission component to the southwest of the main emission region ($2.1\sigma$ post-trial). In order to reject the possibility of an artifact induced by some bias in the background estimation, we performed two tests. First, we repeated the ring background estimation, varying the radius and width of the ring. As a second test, we run a completely independent analysis with the background model using the \texttt{BAccMod}\footnote{\url{https://github.com/mdebony/BAccMod/tree/main}} tool, all yielding consistent results. We note that a soft excess in the same region was reported by VERITAS \citep{2014ApJ...788...78A}, but later observation did not confirm its presence \citep{Abeysekara_2018}. The hot spot is neither coincident with any counterpart shown in Figure~\ref{fig.dragonfly_skymap}, nor there is a nearby Fermi-LAT source in the 4FGL catalog. 

%sst1mpipe_workdir/data_pipe/notebooks/dragonfly_molecular_clouds.ipynb
%Here we can also mention what we see in radio - it is on the edge of one molecular cloud (we can also show an image)

%sst1mpipe_workdir/data_pipe/notebooks/sst1m_significance_map_dragonfly_v074.ipynb
%sst1mpipe_workdir/data_pipe/notebooks/sst1m_dragonfly_edep_morphology_v074.ipynb

In order to test the energy-dependent morphology and to derive the best-fitting coordinates of the VER J2019+368 region in the SST-1M energy band, we created a skymap of the region using \texttt{FoVBackgroundMaker} method of \texttt{gammapy} for the background estimation. We tested the energy-dependent morphology of VER J2019+368 in three logarithmically spaced energy bins between 2.5 TeV and 100 TeV using a symmetric Gaussian spatial model, and a power-law (PL) spectral model with the spectral index fixed at -2 to decrease the degree of freedom of the fit. The best fitting spatial models are shown in Figure~\ref{fig.dragonfly_profile}, right. Testing against the null hypothesis H0 that the morphology of the source is energy-independent using \texttt{EnergyDependentMorphologyEstimator} method of \texttt{gammapy}, we got $\Delta TS = 11.5$ and thus cannot reject H0. Figure~\ref{fig.dragonfly_profile} - right, also shows an asymmetric Gaussian which fits the best the emission in the full range $2.5-100$ TeV, with $\sigma_\mathrm{long} = 0.42^\circ \pm 0.06^\circ, \sigma_\mathrm{lat} = 0.16^\circ \pm 0.03^\circ$, where the extension along the major axis is consistent with the longitudinal profile. The best fitting coordinates of the asymmetric source are $\alpha = 304.81\pm0.07^\circ$, $\delta=36.71\pm0.04^\circ$.

% For extended source skymap with 0.25 deg kernel:
% delta TS = 5.3, 1-100 TeV
% pos: ra 304.81570311 +- 0.06842679, dec 36.75604579 +- 0.02347147
%sigma_long = 0.3560927782798366 deg +- 0.0731438756500747 deg
%sigma_lat = 0.11321929128490009 deg +- 0.04078334063750707 deg
% phi = 74.55702633 deg +- 5.79359106 deg

\subsection{Spectral analysis}
\begin{figure}[!t]
\centering
\begin{tabular}{cc}
\includegraphics[width=0.49\textwidth]{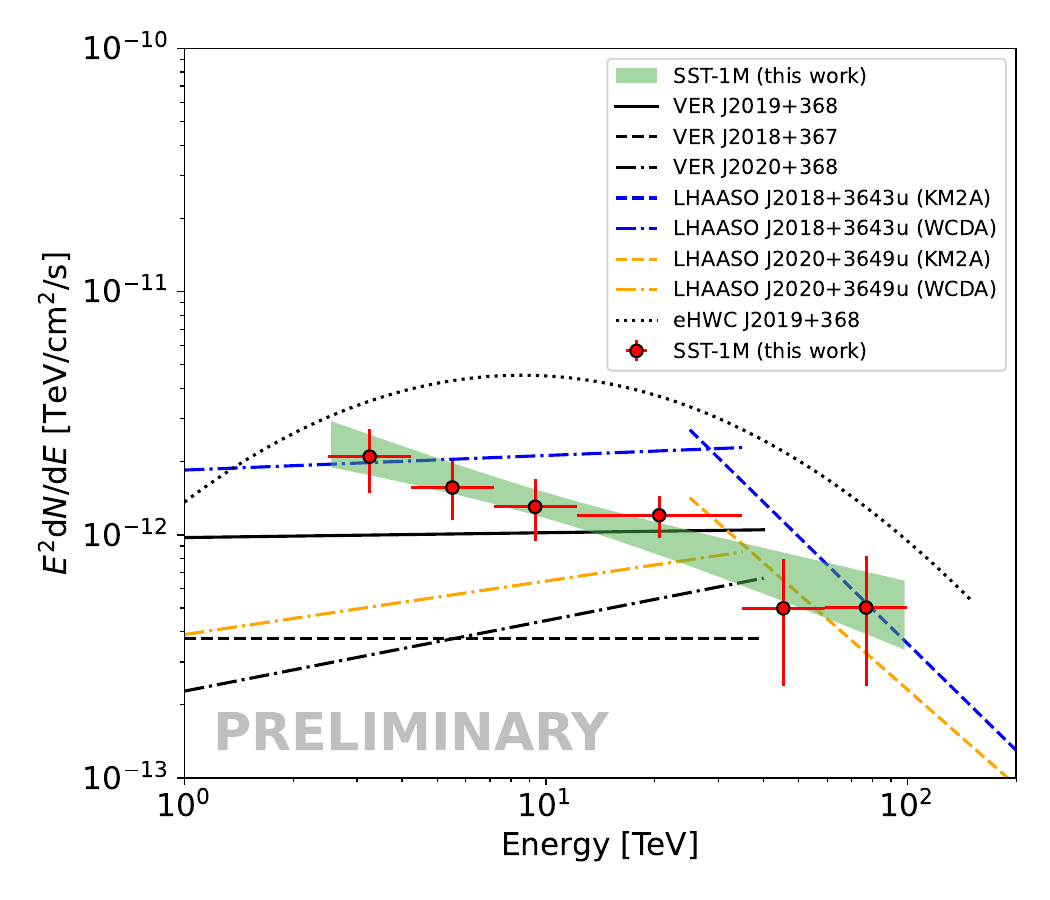} &
\includegraphics[width=0.48\textwidth]{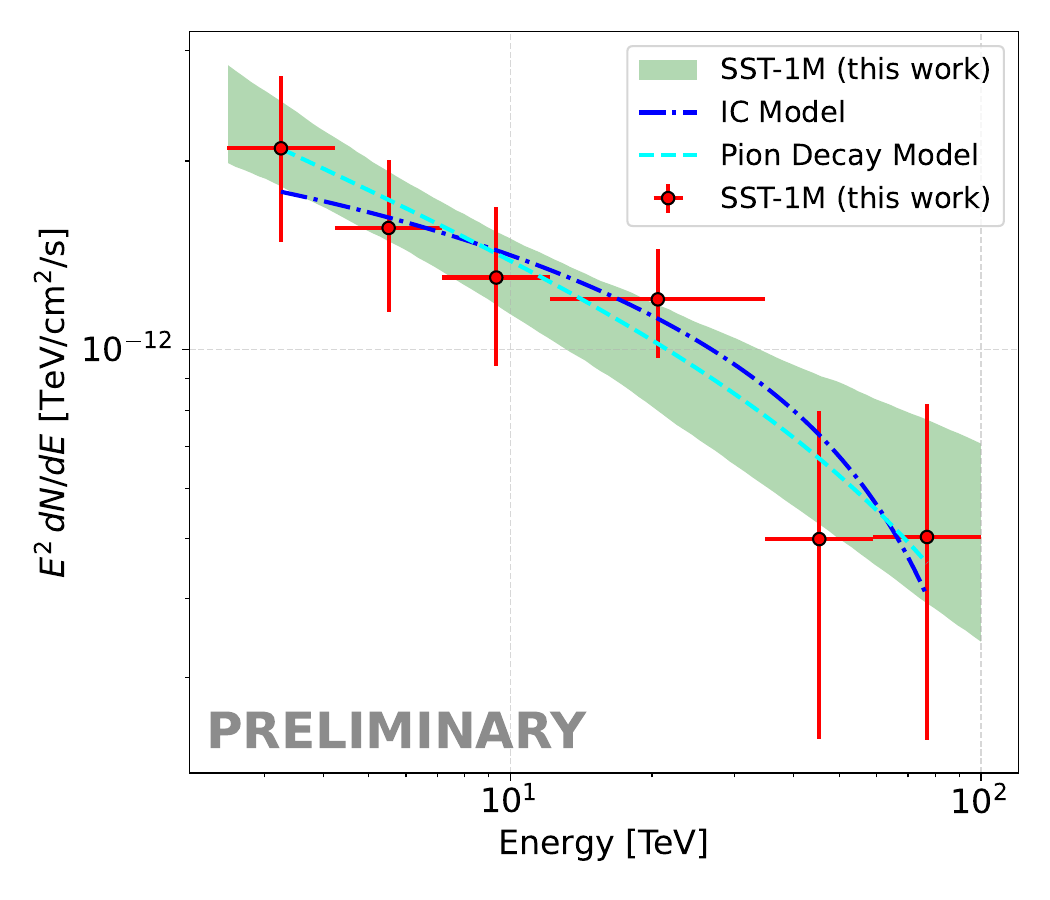} \\
\end{tabular}
\caption{\textit{Left:} SED of the VER J2019+368 region as measured with SST-1M stereoscopic system, compared with results of different observatories (we note, however, that the integration regions are not the same). \textit{Right:} Modeling of the SST-1M SED assuming inverse Compton (IC) and pion decay emission processes, with a simple PL particle distribution.}
\label{fig.sed}
\end{figure}

%sst1mpipe_workdir/data_pipe/notebooks/sst1m_spectrum_dragonfly_v074.ipynb
We performed an independent 1D spectral analysis of the same dataset, adopting the asymmetric Gaussian shape derived in Sec~\ref{sec.morphology} as the signal region. To limit the systematics induced by a possible MC-data disagreement, we rejected all events in the reconstructed energy bins with the effective area $< 5\%$ of the maximum and with the energy bias $> 10\%$. We derived the preliminary spectral energy distribution (SED) of the excess events in the signal region after subtraction of the reflected background, with the background regions selected automatically by \texttt{gammapy} to avoid overlaps, and possible contamination with signal gammas (we used the same exclusion regions as in Sec.~\ref{sec.morphology}). We assumed a Power-Law spectral shape for the differential flux $d\phi / dE$ in a form $d\phi / dE = \phi_0 (E / E_0)^{-\Gamma}$, with the reference energy $E_0 = 7$ TeV. The SED is shown in Figure~\ref{fig.sed}. The best-fitting spectral index $\Gamma=2.44\pm0.13$, and the flux normalization $\phi_0=(3.17\pm0.41) \times 10^{-14} \, \mathrm{cm^{-2} s^{-1} TeV^{-1}}$. After fixing the spectral index $\Gamma$, the flux points were also derived by fitting the flux normalization in 6 energy bins.

%sst1mpipe_workdir/data_pipe/notebooks/sst1m_spectrum_dragonfly_v074.ipynb

We performed a simple one-zone modeling of the SED using \texttt{naima} \citep{naima2015}, assuming that the observed photons originate either from inverse Compton scattering of cosmic microwave background photons or from the decay of pions produced in proton-proton inelastic collisions. In both cases, our data can be well explained with physically plausible parameters (see right panel in Figure~\ref{fig.sed}).

For the inverse Compton scenario, good modeling is achieved using a PL electron distribution injected with a typical spectral index of $\sim 2$ and with synchrotron radiation as the main cooling mechanism, resulting in a particle spectral index of $\sim 3$ and a maximum energy of around 200~TeV. The acceleration of these electrons could be related to the PWN associated with PSR J2021+3651, where efficient acceleration at the wind termination shock might allow electrons to reach such high energies. Similarly, the pion decay scenario is well described by a proton population following a PL with a spectral index of $\sim 2.4$, which could be explained by a combination of Fermi-like acceleration and diffusion of particles in the interstellar medium, with a maximum proton energy of about 1~PeV. These protons could be accelerated by a combination of processes related to both the PWN and the supernova remnant (SNR), or they could have been accelerated during earlier stages of the SNR evolution.

%%%%%%%%%%%%%%
\vspace{-2mm}
\section{Conclusions}
\vspace{-2mm}
We presented preliminary results of the first observations of the Dragonfly region with the SST-1M stereoscopic system, leading to detection of the main emission region, and resolving VER J2019+368 and VER J2016+371 components. At this point, we cannot confirm energy-dependent morphology of the source. Preliminary spectral analysis of the integrated flux from the main extended region VER J2019+368 shows relatively bright PL emission with the spectral index $\Gamma=2.44\pm0.13$ up to 100 TeV. We considered IC and pion decay emission scenarios with electron and proton populations following simple PL, resulting in maximum electron energy of 200 TeV and maximum proton energy of about 1~PeV. Although we considered more complex particle distributions in both cases, simple PL models consistently provided the best fits. However, these results should be interpreted with caution, as more detailed modeling, incorporating a full 3D spectral analysis of our data and multi-wavelength information from the literature, is needed before drawing firm conclusions about the physical processes at work in the region. This will be addressed in a forthcoming study.

\vspace{-2mm}
%%%%%%%%%%%%%%
\section*{Acknowledgements}\scriptsize
\vspace{-2mm}
This publication was created as part of the projects funded in Poland by the Minister of Science based on agreements number 2024/WK/03 and DIR/\-WK/2017/12. The construction, calibration, software control and support for operation of the SST-1M cameras is supported by SNF (grants CRSII2\_141877, 20FL21\_154221, CRSII2\_160830, \_166913, 200021-231799), by the Boninchi Foundation and by the Université de Genève, Faculté de Sciences, Département de Physique Nucléaire et Corpusculaire. The Czech partner institutions acknowledge support of the infrastructure and research projects by Ministry of Education, Youth and Sports of the Czech Republic (MEYS) and the European Union funds (EU), MEYS LM2023047, EU/MEYS CZ.02.01.01/00/22\_008/0004632, CZ.02.01.01/00/22\_010/0008598, Co-funded by the European Union (Physics for Future – Grant Agreement No. 101081515), and Czech Science Foundation, GACR 23-05827S.
\vspace{-3mm}

%%%%%%%%%%%%%%
\bibliographystyle{JHEPe}
{\footnotesize
\bibliography{references}}

%% Full authors list (ONLY FOR COLLABORATIONS)
\clearpage
\section*{Full Authors List: SST-1M Collaboration}\label{authorlist}
\scriptsize
\noindent
C.~Alispach$^1$,
A.~Araudo$^2$,
M.~Balbo$^1$,
V.~Beshley$^3$,
J.~Bla\v{z}ek$^2$,
J.~Borkowski$^4$,
S.~Boula$^5$,
T.~Bulik$^6$,
F.~Cadoux$^`$,
S.~Casanova$^5$,
A.~Christov$^2$,
J.~Chudoba$^2$,
L.~Chytka$^7$,
P.~\v{C}echvala$^2$,
P.~D\v{e}dic$^2$,
D.~della Volpe$^1$,
Y.~Favre$^1$,
M.~Garczarczyk$^8$,
L.~Gibaud$^9$,
T.~Gieras$^5$,
E.~G{\l}owacki$^9$,
P.~Hamal$^7$,
M.~Heller$^1$,
M.~Hrabovsk\'y$^7$,
P.~Jane\v{c}ek$^2$,
M.~Jel\'inek$^{10}$,
V.~J\'ilek$^7$,
J.~Jury\v{s}ek$^2$,
V.~Karas$^{11}$,
B.~Lacave$^1$,
E.~Lyard$^{12}$,
E.~Mach$^5$,
D.~Mand\'at$^2$,
W.~Marek$^5$,
S.~Michal$^7$,
J.~Micha{\l}owski$^5$,
M.~Miro\'n$^9$,
R.~Moderski$^4$,
T.~Montaruli$^1$,
A.~Muraczewski$^4$,
S.~R.~Muthyala$^2$,
A.~L.~Müller$^2$,
A.~Nagai$^1$,
K.~Nalewajski$^5$,
D.~Neise$^{13}$,
J.~Niemiec$^5$,
M.~Niko{\l}ajuk$^9$,
V.~Novotn\'y$^{2,14}$,
M.~Ostrowski$^{15}$,
M.~Palatka$^2$,
M.~Pech$^2$,
M.~Prouza$^2$,
P.~Schovanek$^2$,
V.~Sliusar$^{12}$,
{\L}.~Stawarz$^{15}$,
R.~Sternberger$^8$,
M.~Stodulska$^1$,
J.~\'{S}wierblewski$^5$,
P.~\'{S}wierk$^5$,
J.~\v{S}trobl$^{10}$,
T.~Tavernier$^2$,
P.~Tr\'avn\'i\v{c}ek$^2$,
I.~Troyano Pujadas$^1$,
J.~V\'icha$^2$,
R.~Walter$^{12}$,
K.~Zi{\c e}tara$^{15}$ \\

\noindent
$^1$D\'epartement de Physique Nucl\'eaire, Facult\'e de Sciences, Universit\'e de Gen\`eve, 24 Quai Ernest Ansermet, CH-1205 Gen\`eve, Switzerland.
$^2$FZU - Institute of Physics of the Czech Academy of Sciences, Na Slovance 1999/2, Prague 8, Czech Republic.
$^3$Pidstryhach Institute for Applied Problems of Mechanics and Mathematics, National Academy of Sciences of Ukraine, 3-b Naukova St., 79060, Lviv, Ukraine.
$^4$Nicolaus Copernicus Astronomical Center, Polish Academy of Sciences, ul. Bartycka 18, 00-716 Warsaw, Poland.
$^5$Institute of Nuclear Physics, Polish Academy of Sciences, PL-31342 Krakow, Poland.
$^6$Astronomical Observatory, University of Warsaw, Al. Ujazdowskie 4, 00-478 Warsaw, Poland.
$^7$Palack\'y University Olomouc, Faculty of Science, 17. listopadu 50, Olomouc, Czech Republic.
$^8$Deutsches Elektronen-Synchrotron (DESY) Platanenallee 6, D-15738 Zeuthen, Germany.
$^9$Faculty of Physics, University of Bia{\l}ystok, ul. K. Cio{\l}kowskiego 1L, 15-245 Bia{\l}ystok, Poland.
$^{10}$Astronomical Institute of the Czech Academy of Sciences, Fri\v{c}ova~298, CZ-25165 Ond\v{r}ejov, Czech Republic.
$^{11}$Astronomical Institute of the Czech Academy of Sciences, Bo\v{c}n\'i~II 1401, CZ-14100 Prague, Czech Republic.
$^{12}$D\'epartement d'Astronomie, Facult\'e de Science, Universit\'e de Gen\`eve, Chemin d'Ecogia 16, CH-1290 Versoix, Switzerland.
$^{13}$ETH Zurich, Institute for Particle Physics and Astrophysics, Otto-Stern-Weg 5, 8093 Zurich, Switzerland.
$^{14}$Institute of Particle and Nuclear Physics, Faculty of Mathematics and Physics, Charles University, V Hole\v sovi\v ck\' ach 2, Prague 8, Czech~Republic.
$^{15}$Astronomical Observatory, Jagiellonian University, ul. Orla 171, 30-244 Krakow, Poland.

\end{document}